\documentclass{ws-procs9x6}

\def\beq{\begin{equation}}
\def\eeq{\end{equation}}
\def\beqa{\begin{eqnarray}}
\def\eeqa{\end{eqnarray}}

\def\za{\alpha}
\def\zb{\beta}
\def\ssc{\scriptscriptstyle}
\def\lsim{\mathrel{\raise.3ex\hbox{$<$\kern-.75em\lower1ex\hbox{$\sim$}}} }
\def\gsim{\mathrel{\raise.3ex\hbox{$>$\kern-.75em\lower1ex\hbox{$\sim$}}} }

\begin{document}
\thispagestyle{empty}

\onecolumn

\begin{flushright}
NCU-HEP-k011  \\
Jan 2004
\end{flushright}

\vspace*{.5in}

\begin{center}
{\bf  \boldmath \protect \Large Higgs sector contributions to $\mbox{$\Delta a_\mu$}$ and the 
constraints on two-Higgs-doublet-model with and without SUSY.  }\\
\vspace*{.5in}
{\bf  Otto C.W. Kong}\\[.05in]
{\it Department of Physics, National Central University, \\ Chung-li, TAIWAN 32054 \\
E-mail: otto@phy.ncu.edu.tw}

\vspace*{1.in}
\end{center}
\abstracts{
Interesting contributions to $ \Delta a_\mu$ from a two-Higgs-doublet-model
is coming from a two-loop Barr-Zee diagram for most part of the parameter
space --- a fact that has been overlooked by some Higgs/SUSY experts. 
A definite positive contribution has requirements that go against
precision EW data and other known constraints. For the case without SUSY,
in particular, this is almost enough to kill the two-Higgs-doublet-model 
(II). We will discuss the interplay of all the constraints and their
implications. 
}

\vfill
\noindent --------------- \\
$^\star$ Talk presented  at SUSY 2003 (Jun 4-11), Tucson, AZ USA\\
 --- submission for the proceedings.  
 
\clearpage
\addtocounter{page}{-1}

\title{\bf\boldmath\protect Higgs sector contributions to $\Delta \mbox{\protect $a_\mu$}$ and the 
constraints on two-Higgs-doublet-model with and without SUSY.}

\author{OTTO~C.~W. KONG\footnote{\uppercase{W}ork partially
supported by grant \uppercase{NSC}92-2112-\uppercase{M}-008-044 of the 
\uppercase{N}ational \uppercase{S}cience \uppercase{C}ouncil of \uppercase{T}aiwan.}}

\address{Department of Physics, National Central University,
Chung-li, TAIWAN 32054\\ 
E-mail:  otto@phy.ncu.edu.tw}

\maketitle

\abstracts{
Interesting contributions to  $\Delta a_\mu$ from a two-Higgs-doublet-model
is coming from a two-loop Barr-Zee diagram for most part of the parameter
space --- a fact that has been overlooked by some Higgs/SUSY experts. 
A definite positive contribution has requirements that go against
precision EW data and other known constraints. For the case without SUSY,
in particular, this is almost enough to kill the two-Higgs-doublet-model 
(II). We will discuss the interplay of all the constraints and their
implications. }

\section{\boldmath\protect $\mbox{$\Delta a_\mu$}$ Anomaly and Higgs Contributions} 
This talk is based on the suggestion that there is a disagreement between the
experimentally measured value of the muon anomalous magnetic moment and
that of the SM theoretical value. We are interested in the significance of the 
Higgs sector contributions within the framework of a two-Higgs-doublet-model (2HDM),
with or without SUSY. Apparently, there has been a lack of appreciation for the
fact that the dominating Higgs contributions is coming from a two-loop
Barr-Zee diagram\cite{as13,007}.  We discuss the kind of contributions and
their possible role to the explanation of the  $\Delta a_\mu$ anomaly. 

Most of the specific results used for illustrations here are based on our earlier paper\cite{007}.
In particular, the $\Delta a_\mu$ anomaly number is taken as
\[
\Delta a_\mu \equiv a_\mu^{\rm exp} - a_\mu^{\rm SM} = 
\begin{array}{cc}
(33.9 \pm 11.2) \times 10^{-10} & \quad\mbox{(based on $e^+e^-$ data)} \;,\\
\end{array}
\]
which represents a discrepancy at a $3\,\sigma$  level\cite{smamu}.
It should be noted that if the hadronic vacuum polarization contribution within the 
SM calculations is obtained based on input from  $\tau$ data, instead of the
$e^+e^-$ data, the discrepancy would be reduced\cite{smamu}.
\\
\parbox{2.5in}{\ \ Our focus here is the Higgs sector contributions. A 1-loop diagram
has a contribution too small to explain the discrepancy for $m_\phi > 10$ GeV\cite{1loop}. 
However, at the 2-loop level, there is a (photon) Barr-Zee diagram (as shown to the right) 
with contribution easily {dominates} over the 1-loop result for $m_\phi > 3$ GeV. The 
diagram may have enhancement from a large  $\tan\!\beta$. We have
}
\begin{minipage}[h]{1in}
\vspace*{1.8in}
\includegraphics{as13-1.ps}
\end{minipage}
\vspace*{-.5in}
\beq
\Delta a_\mu^{\phi} =
\frac{N_{\!c}^f \, \alpha_{\mbox{\tiny em}}}{4\pi^3 \,v^2} 
{m_\mu^2}\;
{\mathcal Q}_f^2 \left[ A_\mu \, A_f \,
g\!\!\left( \frac{m_f^2}{m_{\phi}^2} \right)
- \lambda_\mu \, \lambda_f \,
f\!\!\left( \frac{m_f^2}{m_\phi^2} \right) \right] \;,
\vspace*{-.3in}
\eeq
where $\lambda_f$ and $A_f$ represent the effective scalar and pseudoscalar
couplings of a fermion $f$ to the Higgs state, with loop functions
$f(z)={1\over 2} z \int^1_0 \! dx \; 
\frac{1-2x(1-x)}{x(1-x)-z} \ln\frac{x(1-x)}{z}$ and
$g(z)={1\over 2} z \int^1_0 \!  dx \; \frac{1}{x(1-x)-z} \ln\frac{x(1-x)}{z}$. 
\\

The most interesting point to note is that the 2-loop contribution has, in general, an
opposite sign to the 1-loop result. It is negative for a real scalar, but positive for a
pseudoscalar. With a minimally extended Higgs sector, a 2HDM has two real scalars
($h$ and$H$) and a pseudoscalar ($A$). If the latter contribution dominates, there
is a chance that the Higgs sector contributions can account for the $\Delta a_\mu$
anomaly. We illustrate in the plot below the result from the contribution of a single
pseudoscalar (1-loop + photon Barr-Zee, with SM fermions) with the variations of
$m_{\ssc A}$ and $\tan\!\beta$. In the case of a concrete model, the cancellation
effect from the negative scalar contributions has also to be taken into account.
\\
\parbox{2.in}{
\section{On the Two-Higgs-Doublet-Model} 
The 2HDM II is the most appealing Higgs sector extension, and a natural component
of the supersymmetric SM. Neglecting the very small admissible CP violation,
we have the following results on the relative couplings for $t$, $b$, and $\tau$, respectively :
}
\begin{minipage}[h]{2.in}
\includegraphics{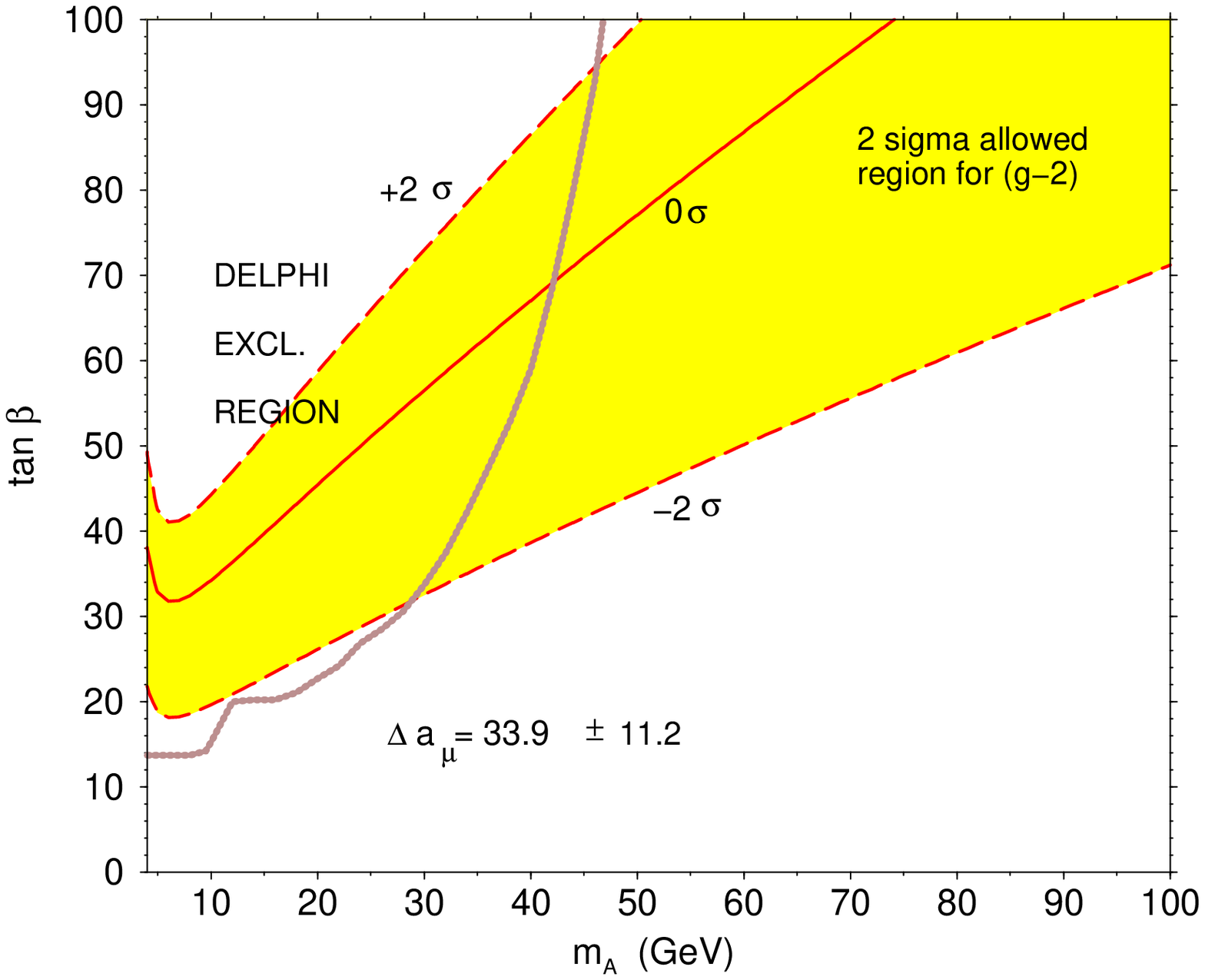}
\end{minipage}

\eject
\noindent
\parbox{2.1in}{
\small
\beqa
 h \;(\lambda_{f}) & :&  \quad  \frac{\cos\!\za}{\sin\!\zb}
    \quad  -\frac{\sin\!\za}{{\cos\!\zb}}  \quad  -\frac{\sin\!\za}{{\cos\!\zb}}
\nonumber \\
H\; (\lambda_{f})  & :&  \quad  \frac{\sin\!\za}{\sin\!\zb}
     \qquad  \;\frac{\cos\!\za}{{\cos\!\zb}}  \qquad \;\frac{\cos\!\za}{{\cos\!\zb}}
\nonumber \\
{A} \;(A_{f})  & :&  \quad {\cot\!\zb} \qquad \;\;{\tan\!\zb} \qquad \;\,{\tan\!\zb}
 \;. \quad\nonumber 
\eeqa
\normalsize
\ \ From the above, it is clear that for the Higgs sector contributions to account for
any substantial part of the $\Delta a_\mu$ anomaly, a light pseudoscalar together
with heavy scalars and a relatively large $\tan\!\zb$ would be required. The
condition $m_{\ssc A} \!\!< \!\!m_{\ssc h}$ is not admissible in the SUSY case. 
However, one is still left with the 
question if the Higgs sector contributions could have a significant role to play, may
be giving a substantial negative overall contribution to shift the parameter
space solution region from that of the naive 1-loop considerations. While the 
possible role of the Barr-Zee diagrams, here extended to includes the ones with
sfermions running in the upper loop, in the SUSY case for the study of EDM is
well documented, the corresponding situation of the magentic moments is largely
overlooked. Studies of fitting $\Delta a_\mu$ focused only on the 1-loop chargino 
and neutralino contributions. Fortunately, we obtained a definite negative
result \cite{007}, for a generic choice of SUSY parameters. 

\ \ For the case without SUSY,  while a light pseudosclar is admissible, the fit
the required $\Delta a_\mu$ numerical, a substantial splitting between 
$m_{\scriptscriptstyle A}$  and $m_h\;( < m_{\scriptscriptstyle H}\,)$ is needed\cite{as13}.
\ In fact, \ the $\Delta a_\mu$ \ anomaly imposes a}
\begin{minipage}[h]{2.in}
\vspace*{2.5in}
\includegraphics{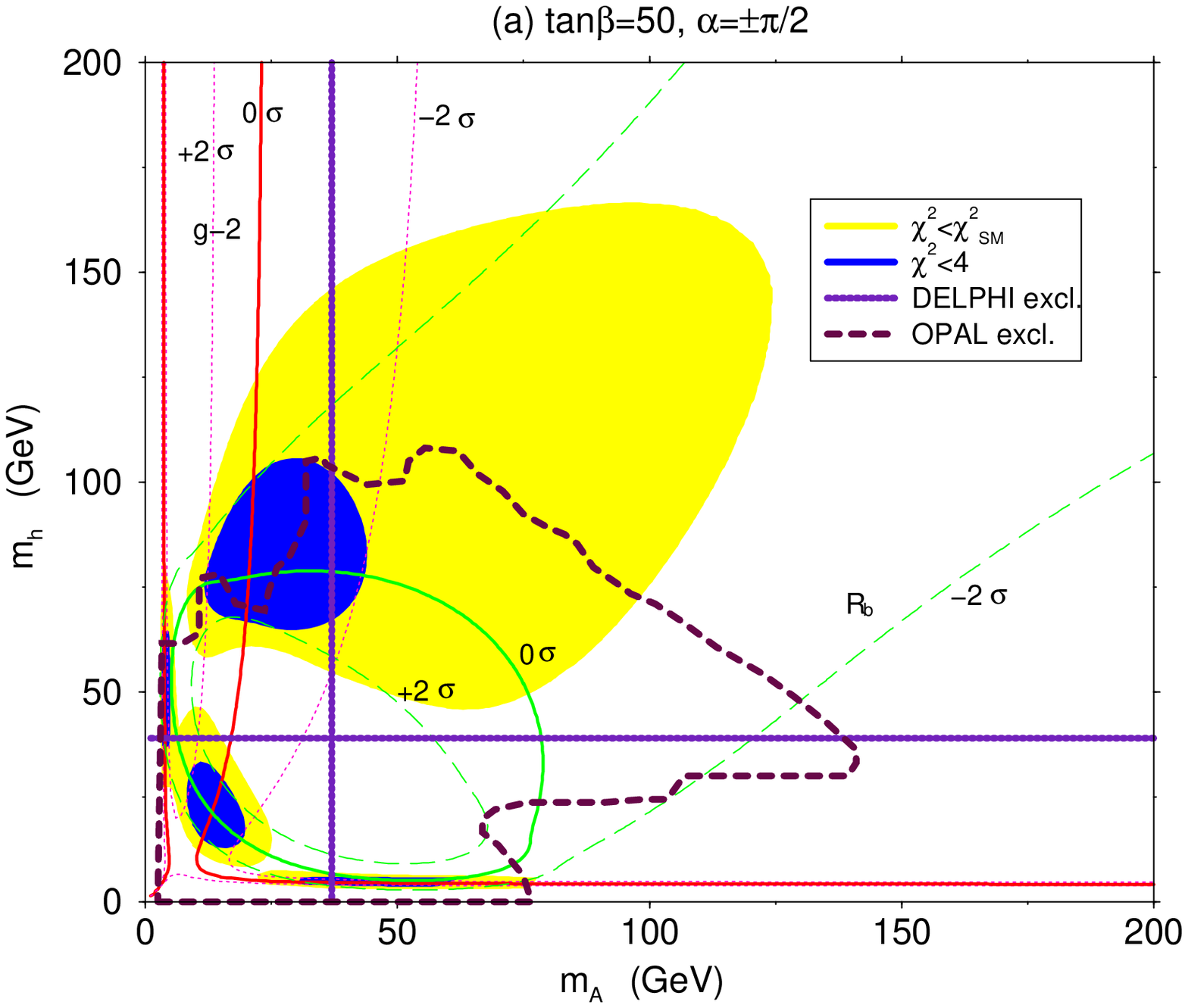}

\vspace*{2.4in}
\includegraphics{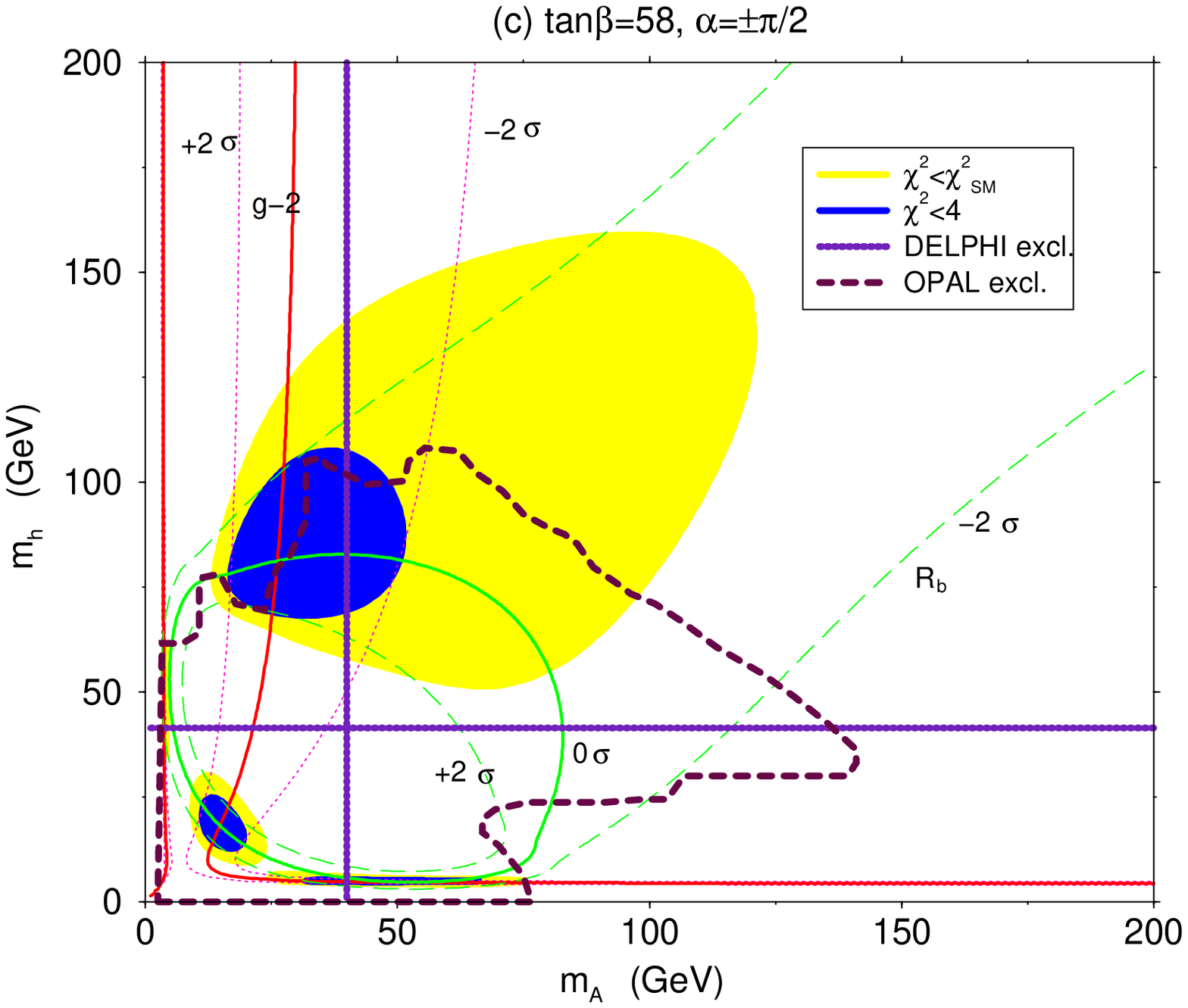}

\vspace*{2.4in}
\includegraphics{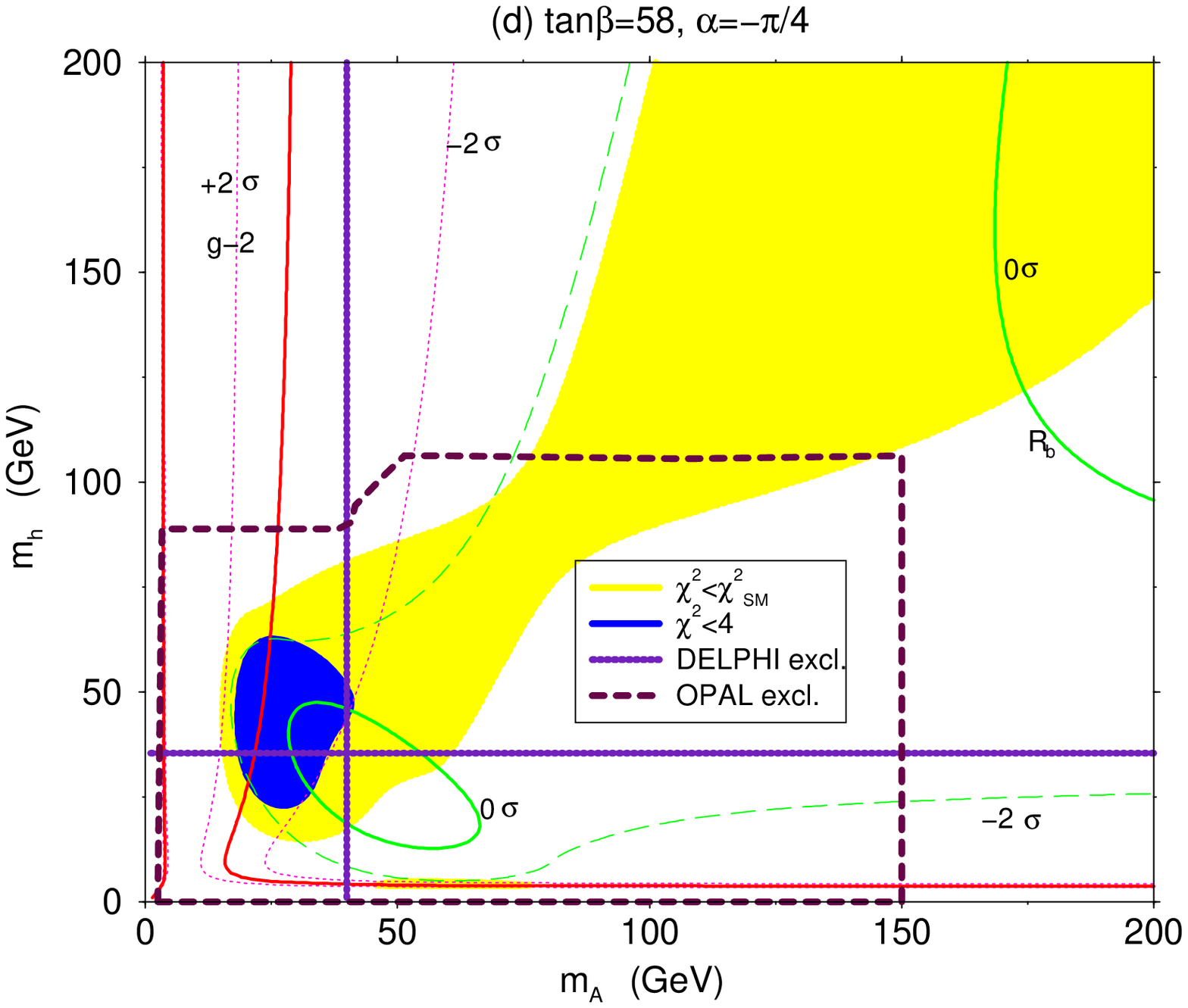}
\vspace*{.1in}
{\small \hspace*{.2in}\hbox{Plots of all constraints on 2HDM.}}
\end{minipage}

\eject
\noindent
very stringent constraint on the 
model in a way largely complementary to the precision EW, and other known, 
constraints. The interplay of all these is very interesting. For example, taking the 
$3.3\,\sigma$ $\Delta a_\mu$ together with the $R_b$ constraint with a limit
$\chi^2 < 4$, the 2HDM would  largely be ruled out.

\section{Putting Together the Other Constraints}
We have performed a comprehensive study of the overall Higgs sector contributions, 
together with other available constraints on the model\cite{007}. We illustrate a few plots 
from our result on the previous page. The dark color shaped regions represent
solutions to  $\chi^2 < 4$ for $\Delta a_\mu$ and the $R_b$ constraints combined.
The light color shaped regions have   $\chi^2 < 10.3$, the SM value. Here, the $R_b$
constraint used is given by 
$
{\Delta R_b} \equiv R_b^{\rm exp} - R_b^{\rm SM} = 
{ 0.000692 \pm 0.00065}  \;.
$
We have, for each plot, pick a choice of values for $\tan\!\zb$ and the scalar Higgs
mixing angle $\za$. The heavy scalar $H$ is assumed to be heavy enough for its 
effects to be neglected, while the charged Higgs mass is set at $500\,\mbox{GeV}$.
The charged Higgs contributes, through the
charged current interaction with CKM mixings, strongly to $b\to s\,\gamma$.
Our choice of $m_{\ssc H^+}$ is then about the lowest admissible value.
Our careful analysis of the constraint from the $\rho$-parameter\cite{007}
illustrated that the Higgs masses are forced into a very fine-tuned 
relation, which, for the case of $m_{\ssc A}\!\!<\!\!m_{\ssc h}\ll m_{\ssc H^+}$,
requires $m_{\ssc H}$ to be at least a few times $m_{\ssc H^+}$. Hence, it
justifies our neglecting $H$ contributions (to $R_b$ and $a_\mu$). 
Finally, we also lput in the experimental bounds from OPAL and DELPHI\cite{e}.
The first two plots shown represent what is close to the best case scenario. 
Any possible surviving region in the parameter space would have an $\za$
value around ${-3\pi/8}$ to  ${-\pi/2}$, and a value of $\tan\!\zb$ that
is uncomfortably large.

{\it This write-up is done while the author is visiting as the Korea Institute for Advanced Study.
The institute is to be thanked for the great hospitality. The author is in debt to K.~Cheung,
from the collaboration results with whom that the presentation is based.}

\end{document}